**Title**:

**Rapid AI Development Cycle for the Coronavirus (COVID-19) Pandemic: Initial Results for Automated Detection & Patient Monitoring using Deep Learning CT Image Analysis**

**Article Type**:

Original Research

**Authors:**

Ophir Gozes (1)(+),  Ma'ayan Frid-Adar (1)(+), Hayit Greenspan, PhD (*)(1)(2), Patrick D. Browning, MD (1), Huangqi Zhang, MD (3), Wenbin Ji, MD (3),  Adam Bernheim, MD (4), and Eliot Siegel, MD (5)

  (1) RADLogics
  (2) Tel-Aviv University
  (3) Affiliated Taizhou Hospital of Wenzhou Medical University, 150 Ximen Street, Linhai, 317000, Zhejiang Province, China
  (4) Mount Sinai Hospital, New York, NY
  (5) University of Maryland School of Medicine, Baltimore, MD
    *(+) Equal Contributers*
    *(*) Corresponding Author*

**Summary Statement**:

Rapidly developed AI-based automated CT image analysis tools can achieve high accuracy in detection of Coronavirus positive patients as well as quantification of disease burden.

**Key Results:**

- Utilizing the deep-learning image analysis system developed, we achieved classification results for Coronavirus vs Non-coronavirus cases per thoracic CT studies of 0.996 AUC (95%CI: 0.989-1.00) on Chinese control and infected patients. Possible working point: 98.2% sensitivity, 92.2% specificity.

- For Coronavirus patients the system outputs quantitative opacity measurements and a visualization of the larger opacities in a slice-based "heat map" or a 3D volume display. A suggested "Corona score" measures the progression of patients over time.

## List of abbreviations:

- AI—Artificial Intelligence
- DL—Deep Learning
- CT—Computer Tomography
- COVID-19—Corona Virus 2019

# Abstract

- **Background:**

The coronavirus infection surprised the world with its rapid spread and has had a major impact on the lives of billions of people. Non-contrast thoracic CT has been shown to be an effective tool in detection, quantification and follow-up of disease. Deep learning algorithms can be developed to assist in analyzing potentially large numbers of thoracic CT exams.

- **Purpose:**

To develop AI-based automated CT image analysis tools for detection, quantification, and tracking of Coronavirus and demonstrate that they can differentiate coronavirus patients from those who do not have the disease.

- **Materials and Methods:**

Multiple international datasets, including from Chinese disease-infected areas were included. We present a system that utilizes robust 2D and 3D deep learning models, modifying and adapting existing AI models and combining them with clinical understanding.
We conducted multiple retrospective experiments to analyze the performance of the system in the detection of suspected COVID-19 thoracic CT features and to evaluate evolution of the disease in each patient over time using a 3D volume review, generating a "Corona score". The study includes a testing set of 157 international patients (China and U.S).

- **Results***:*

Classification results for Coronavirus vs Non-coronavirus cases per thoracic CT studies were 0.996 AUC (95%CI: 0.989-1.00) ; on datasets of Chinese control and infected patients. Possible working point: 98.2% sensitivity, 92.2% specificity.
For time analysis of Coronavirus patients, the system output enables quantitative measurements for smaller opacities (volume, diameter) and visualization of the larger opacities in a slice-based "heat map" or a 3D volume display. Our suggested "Corona score" measures the progression of disease over time.

- **Conclusion:**

This initial study, which is currently being expanded to a larger population, demonstrated that rapidly developed AI-based image analysis can achieve high accuracy in detection of Coronavirus as well as quantification and tracking of disease burden.

# I. Introduction

The coronavirus infection, COVID-19 has surprised the world with its rapid spread, potential virulence, with potential profound overall impact on the lives of billions of people from both a safety and an economic perspective. As of this writing, there are approximately 93,158 confirmed cases of which 80,270 are in "Mainland China" with 3,198 deaths, a mortality rate of 3.4% [1].

The sensitivity of the current diagnostic gold standard at the initial presentation of the disease has been called into question. Fang et al [2] compared the sensitivity of non-contrast chest CT with reverse transcription-polymerase chain reaction (RT-PCR) which detects viral nucleic acid and is the current reference standard in the detection of COVID-19. Their study looked at 51 patients who had a history of travel or residence in endemic areas and fever or acute respiratory symptoms of unknown cause. The patients underwent initial and repeat RT-PCR testing. Their gold standard was an eventual confirmed diagnosis of COVID-19 infection by serial RT-PCR testing. The authors suggested a sensitivity of non-contrast chest CT for detection of COVID-19 infection of 98% compared to initial RT-PCR sensitivity (results of the first RT_RPR test) of 71%. Cases highlighted in their paper demonstrated either diffuse or focal ground-glass opacities. This lack of sensitivity on initial RT-PCR testing was also described in another study by Xie et al [3] who reported that 3% of 167 patients had negative RT-PCR for the virus despite chest CT findings typical of viral pneumonia suggesting the use of chest CT to decrease false negative lab studies. Bernheim et al [4] examined 121 chest CT studies from four centers in China that were obtained in the early, intermediate and late stages of infection. They also described ground glass opacities as characteristic of the disease, particularly bilateral and peripheral ground-class and consolidative pulmonary opacities. They noted greater severity of disease with increasing time from onset of symptoms and they described later signs of disease to include "greater lung involvement, linear opacities, "crazy-paving" pattern and the "reverse halo" sign. There was bilateral lung involvement in 28% of early patients, 76% of intermediate patients and 88% of patients in late stages (6-12 days) of disease.

Once a decision has been made to use thoracic CT as these recent studies suggest for patient diagnosis or screening, a need immediately rises to rapidly evaluate potentially very large numbers of imaging studies. AI technology, in particular deep learning image analysis tools, can potentially be developed to support radiologists in the triage, quantification, trend analysis of the data. AI solutions have the potential to analyze multiple cases in parallel to detect whether chest CT reveals any abnormalities in the lung. If the software suggests a significantly increased likelihood of disease, the case can be flagged for further review by a radiologist or clinician for possible treatment/quarantine. Such systems, or variations thereof, once verified and tested – can become key contributors in the detection and control of patients with the virus.

In a manner analogous to the way in which COVID-19 represents a new strain of coronavirus not previously found in humans and presumably representing a mutation of other coronaviruses, an AI algorithm can be rapidly created from one or more algorithms that perform a similar task. This is in contrast to the standard way of developing a DL algorithm, entailing several phases: I. Data-collection phase in which a large amount of data samples need to be collected from predefined categories; expert annotations are needed for ground-truthing the data; II. Training phase in which the collected data is used to train network models. Each category needs to be represented well enough so that the training can generalize to new cases that will be seen by the network in the testing phase. In this learning phase, the large number of network parameters (typically on the order of millions) are automatically generated; III. Testing phase in which an additional set of cases not used in training is presented to the network and the output of the network is tested statistically to determine its success of categorization.

In the case of a new disease, such as the coronavirus, datasets are just now being identified and annotated. There are very limited data sources as well as limited expertise in labeling the data specific to this new strain of the virus in humans. Accordingly, it is not clear that there are enough examples to achieve clinically meaningful learning at this early stage of data collection despite the increasingly critical importance of this software, especially given fears of a pandemic. It is our hypothesis that AI-based tools can be rapidly developed leveraging the ability to modify and adapt existing AI models and combine them with initial clinical understanding to address the new challenges and new category of COVID-19. Our goal is to develop deep-learning based automated CT image analysis tools and demonstrate that they can enable differentiation of coronavirus patients from those who do not have the disease to provide support in the detection, measurements, and tracking of disease progression.

## II. Methods

The system we propose receives thoracic CT images and flags cases suspected with COVID-19 features. In addition, for cases classified as positive, the system outputs a lung abnormality localization map and measurements. Figure 1 shows a block diagram of the developed system. The system is comprised of several components and analyzes the CT case at two distinct levels: *Subsystem A*: 3D analysis of the case volume for nodules and focal opacities using existing, previously developed algorithms and *Subsystem B*: newly developed 2D analysis of each slice of the case to detect and localize larger-sized diffuse opacities including ground glass infiltrates which have been clinically described as representative of the coronavirus.

For Subsystem A we use commercial off-the-shelf software that detects nodules and small opacities within a 3D lung volume (RADLogics Inc., Boston [5]). This software was developed as a solution for lung pathology detection and provides quantitative measurements (including volumetric measurements, axial measurements (RECIST), HU values, calcification detection and texture characterization for solid vs sub-solid vs GG). Because ground-glass opacities (GGO) have emerged in recent studies as one of the key features for COVID-19, we hypothesized that the existing software can detect smaller-sized focal GG opacities within a case. An example of this can be seen in Figure 2(A): two Coronavirus cases are shown in which the opacities are relatively subtle. In addition to the detection of abnormalities, measurements and localization are provided. In each case, the software detected a single focal opacity (outlined by a red bounding-box). An image of the detected opacity is shown, along with its segmentation. Finally, a list of lesion features is automatically generated and provided.

Since current lung pathology detection solutions were built with a specific focus on the nodule detection task, they cannot be relied upon for detecting more *diffuse global GG opacities.* To address the additional disease-driven opacities, we proposed a data-driven solution on a per-slice basis, as shown in Figure 1, Subsystem B. Working in the 2D (slice) space has several advantages for Deep-Learning based algorithms, in limited data scenarios. These include an increase in training samples (with many slices per single case), using pre-trained networks that are common in the 2D space, and easier annotation for segmentation purposes.

In our solution (B), the first step is the *Lung Crop stage*: we extract the lung region of interest (ROI) using a lung segmentation module. The U-net architecture for image segmentation [6,7] was trained using 6,150 CT slices of cases with lung abnormalities and their corresponding lung masks which were taken from a U.S based hospital (Table I: Dataset-6). The segmentation step enables the removal of image portions that are not relevant for the detection of within-lung disease making the learning process of the next step easier. In the following step, we focus on *Detecting Coronavirus related abnormalities*: We use a Resnet-50 - 2D deep convolutional neural network architecture [8]; The network is 50 layers deep and can classify images into 1000 categories. The network was pre-trained on more than a million images from the ImageNet database [9]. As commonly done in the medical imaging field, we further train the network parameters (fine-tune) to solve the problem at hand: suspected COVID-19 cases from several Chinese hospitals are used (Table I: Dataset-1). The cases were annotated per slice as normal (n=1036) vs abnormal (n=829). To overcome the limited amount of cases, we employ data augmentation techniques (image rotations, horizontal flips and cropping). In a follow-up *abnormality localization step*, given a new slice classified as positive, we extract "network-activation maps" which correspond to the areas most contributing to the network's decision. This is performed using the Grad-cam technique for producing visual explanations for network decisions [10].

Example results of four COVID-19 slices that we classified as abnormal by the network can be seen in Figure 2(B). On top, the CT image is shown. In the bottom row, corresponding colored maps are provided.

In red we see the strongest network output while blue is the weakest. We note the maps align well with the diffused opacities, providing a strong indication that the network managed to learn important characteristics associated with COVID-19 manifestations.

To mark a case as COVID-19 positive, we calculate the ratio of positive detected slices out of the total slices of the lung (*positive ratio*). A positive case-decision is made if the *positive ratio* exceeds a pre-defined threshold.

To provide a complete review of the case, we combine the output of Subsystem A - 3D analysis and Subsystem B - 2D slice-level. In Figure 3 we see a case of Coronavirus and the combined output findings map from the proposed system. We can see the nodular and focal diffuse opacity detections in green and the larger opacity detection in red. The two subsystems complement and, in some locations reinforce each other.

In addition to the visualization, the system automatically extracts several outputs of interest, including per slice localization of opacities (2D), and a 3D volumetric presentation of the opacities throughout the lungs. We also propose a *Corona score* which is a volumetric measurement of the opacities burden. The corona score is computed by a volumetric summation of the network-activation maps. The score is robust to slice thickness and pixel spacing as it includes pixel volume. For patient-specific monitoring of disease progression, we suggest the *Relative Corona score* in which we normalize the corona score by the score computed at the first time point.

## III. Results

A set of experiments is conducted next to demonstrate the performance of the automated analysis.

*I. Classification:*

We start with an evaluation of the ability to detect *slice-level* Coronavirus. The performance of this step is crucial for obtaining overall case wise detection. For the validation step, we used 10% of the slices from the development dataset comprised of cases from the Chinese population (Table I: Dataset-1). The split was patient wise and there is no overlap with the slices used for training. A total of 270 slices were analyzed: 150 normal slices and 120 COVID-19 suspected slices. Area Under Curve (AUC) was 0.994 with 94% sensitivity and 98% specificity (at threshold 0.5).

We proceed with an evaluation of the ability to detect at the *case-level* for Coronavirus vs non-Corona virus patients. First, we evaluate the classification solely on the Chinese population. For Coronavirus patients, we use a collection of 56 patients with confirmed COVID-19 diagnosis (Table I: Dataset-2) and for non-Coronavirus patients, we use a dataset of 51 Chinese patients from multiple institutions in China (Table I: Dataset-3). Using the *positive ratio* as a decision feature, we were able to achieve an AUC of 0.996 (95%CI: 0.989-1.00). The power of using AI for screening comes from the ability to tune the operating point to support various clinical scenarios. Basing the case level decision on the *positive ratio*: A threshold of 1.1% (percent of positive detected slice to lung slices) yields a case level sensitivity of 98.2% with 92.2% specificity. A threshold of 1.9% yields a sensitivity of 96.4% and specificity of 98%.

To explore the relevance of the solution to a non-Chinese population, we perform a second analysis with the same 56 Chinese patients as COVID-19 positive group, using a U.S source for 49 non-Coronavirus patients (Table I: Dataset-4,5). For this population, we achieve AUC of 0.996 (95%CI: 0.989-1.00), sensitivity of 98.2% and specificity of 91.8% at a positive ratio of 1% and sensitivity of 94.6% and specificity of 98% at a threshold of 4.3%.

Both ROC curves are displayed in Figure 4A. In the ROC calculation, for Coronavirus patients that include multiple time points, the analysis was performed using the first time point in the series.

II. *Evaluation over time:*

In our final experiment, we evaluate patients that were imaged in time points for whom the first CT scan was obtained 1-4 days following the first signs of the virus (fever, cough). In the first example patient, we review a case with a single focal opacity and present volumetric measurements over time. The second case involves patient with multiple opacities and shows an overview of the patient recovery process with its corresponding Corona score over time.

Figure 5 shows tracking over time of a specific opacity in a Coronavirus patient (red box). The four CT scans shown across different time points show a reduction in the lesion size. For each time point, the volume measurement and average axial diameter measurement are displayed at the bottom. The first scan (Jan 26) was taken 1-4 days following symptoms of COVID-19, with a measured volume of 9.9 cm3. The second scan (Jan 30) was taken 5-8 days after the first symptoms and shows a higher volume measurement of 19.9 cm3. The third and fourth scans (Feb 4 and Feb 11) taken 10-13 days and 17-20 days, after first symptoms, respectively, show a decrease in the lesion volume suggesting resolution of disease.

Figure 6 presents an entire case review of a second Coronavirus patient who underwent three CT scans throughout the disease. The 3D lung volume is displayed in blue with the proposed system opacity detection and localization output displayed in red and green. In addition to the representation of the case, the Corona score is calculated for each time point to give a quantitative representation of the disease: The patient's first thoracic CT was obtained Jan 27 (1-4 days after symptoms of COVID-19 were present). Opacities, when present in this patient, appear to more frequently in the mid- and upper-lungs, and less frequently in the lower lobes. The Corona score at this time point is 191.5 cm3. The second CT scan was taken 4 days later, on Jan 31. We can visually detect a reduction in the opacities within the lung volume with Corona score of 97.1 cm3. In comparison to the first scan, we can see and quantitate a reduction of 49% in overall opacity-burden. The final CT scan was taken 15 days after the second scan; here no opacities are present (with Corona score: 0); indicating full thoracic CT finding based resolution of disease.

Figure 4.B displays quantitative assessment scores for 18 COVID-19 patients that were monitored for 30 Days. On the left, we see a plot of the corona score. In this plot, we can assess the relative severity of coronavirus among the patients. A plot of the Relative Corona Score is shown on the right. Here, different courses of the disease can be identified.

# IV. Discussion

In this initial exploratory work, we show the capabilities of AI to assist in the efforts to accurately detect and track the progression or resolution of the Coronavirus. This is the first report to our knowledge in the literature of software specifically developed to detect, characterize and track the progression of COVID-19.

Rapidly developed AI-based automated CT image analysis tools can achieve high accuracy in the detection of Coronavirus positive patients as well as quantification of disease burden. Utilizing the deep-learning image analysis system developed, we achieved classification results for Coronavirus vs Non-coronavirus cases per thoracic CT studies of 0.996 AUC (95%CI: 0.989-1.00) on datasets of Chinese control and infected patients. Two possible working points are: 98.2% sensitivity, 92.2% specificity (high sensitivity point); 96.4% sensitivity, 98% specificity (high specificity point). For Coronavirus patients the system outputs quantitative opacity measurements and a visualization of the larger opacities in a slice-based "heat map" or a 3D volume display. A suggested "Corona score" measures the progression of patients over time.

A consistent and reproducible method for rapid evaluation of high volumes of screening or diagnostic thoracic CT studies using AI can assist in this crisis in several ways: Highly accurate systems can reliably exclude CTs which are negative for findings associated with the corona virus. This decreases the volume of cases passing through to the radiologist without overlooking positive cases. Progression and regression of findings could be monitored more quantitatively and consistently. This would allow a greater volume of patients being screened for Coronavirus, with earlier and more rapid detection of positive cases, which could lead to more effective identification and containment of early cases.

As illustrated above, using standard machine learning techniques and innovative AI applications, in combination with an established pulmonary CT detection platform, an effective tool can be utilized for the screening and early detection of patients who may have contracted the COVID-19 pathogen. In individual patients who have contracted the virus and have the pulmonary abnormalities associated with it, the same methodologies can be used to accurately and more rapidly assess disease progression and guide therapy and patient management.

# Figure Legends

**Figure 1:**  System Block Diagram

**Figure 2A:** Subsystem A: Focal opacities detection and measurements.

**Figure 2B:** Subsystem B: Slices classified as positive for Coronavirus abnormalities and their corresponding "heatmap".

**Figure 3:**  Patient case visualization. Left: Coronal view; Right: Automatically generated 3D volume map of focal opacities (green) and larger diffuse opacities (red).

**Figure 4A**: ROC curves for the task of COVID-19 detection using *positive ratio* feature.

**Left:** Chinese Population
**Right:** Coronavirus patients from China; Non-Coronavirus patients from US

**Figure 4B**: Tracking of patient's disease progression over time using *Corona Score* (Left) and *Relative Corona Score* (Right). Day 0 corresponds to 1-4 days following first signs of the virus.

**Figure 5**:  Multi time point tracking of patient disease progression

**Figure 6:**  Corona score for patient disease progression monitoring

**TABLE I: Datasets**

| | |
|---|---|
| Dataset 1: Development Dataset<br><br>Source: Chainz [11] | 50 abnormal thoracic CT scans (slice thickness, {5,7,8,9,10}mm) from China of patients that were diagnosed by a radiologist as suspicious for COVID-19 (from Jan-Feb 2020). The cases were extracted by querying a cloud PACS system for cases that were referred for laboratory testing following the scan.<br>Cases were annotated for each slice as normal (n=1036) vs abnormal (n=829) |
| Dataset 2: Testing Dataset<br><br>Source: Hospital in Wenzhou, China | 56 patients with confirmed diagnosis of COVID-19 infection by RT-PCR testing. Each patient had a chest CT scan (slice thickness, {1,1.5,5}mm) at one or more time points (between 1 and 5) |
| Dataset 3: Testing Dataset<br><br>Source: Chainz [11] | 51 normal thoracic CT scans (slice thickness, {1,1.3,4,5,9,10}mm) of different patients without any abnormal lung findings in the radiologist's report |
| Dataset 4: Testing Dataset<br><br>Source: El-Camino Hospital (CA) | 30 normal lung chest CT scans (slice thickness, {0.8,1,1.5,2.5,3}mm) collected between 2014-2016 |
| Dataset 5: Testing Dataset<br><br>Source: LIDC [12] | 19 normal chest CT scans (slice thickness, {1.3,2.5}mm) of diagnostic and lung cancer screening patients |
| Dataset 6: Lung segmentation Development<br><br>Sources: El-Camino Hospital (CA) | 6,150 CT slices of cases with lung abnormalities and their corresponding lung masks |

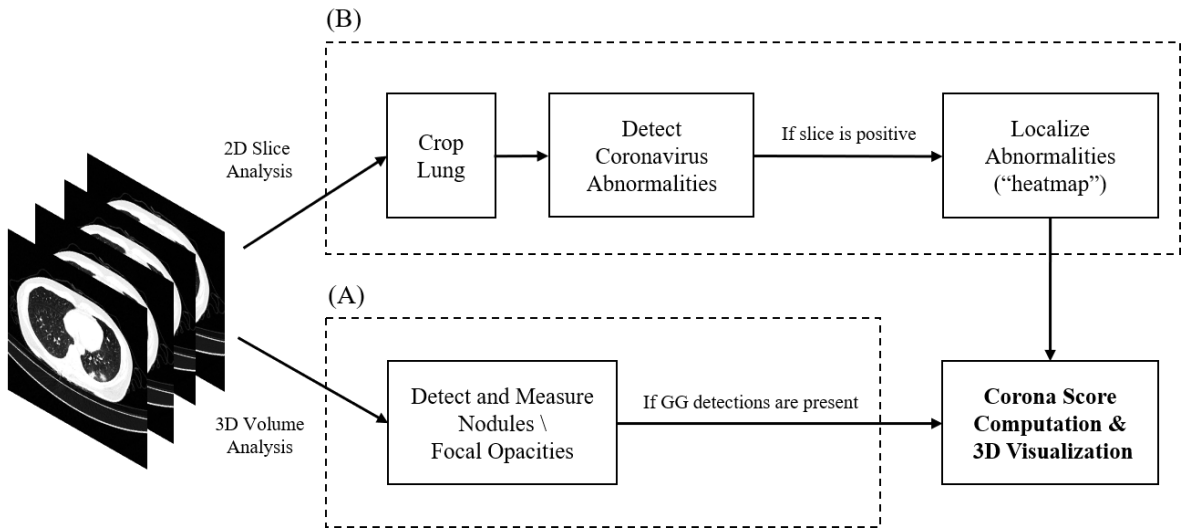

**Figure 1:** System Block Diagram

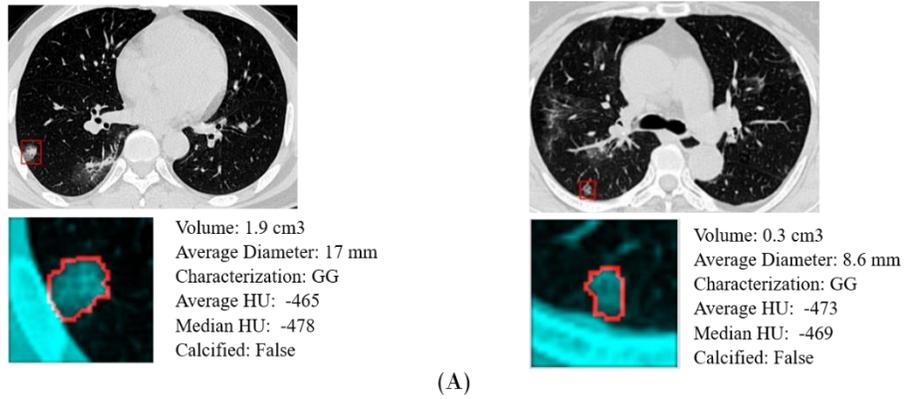

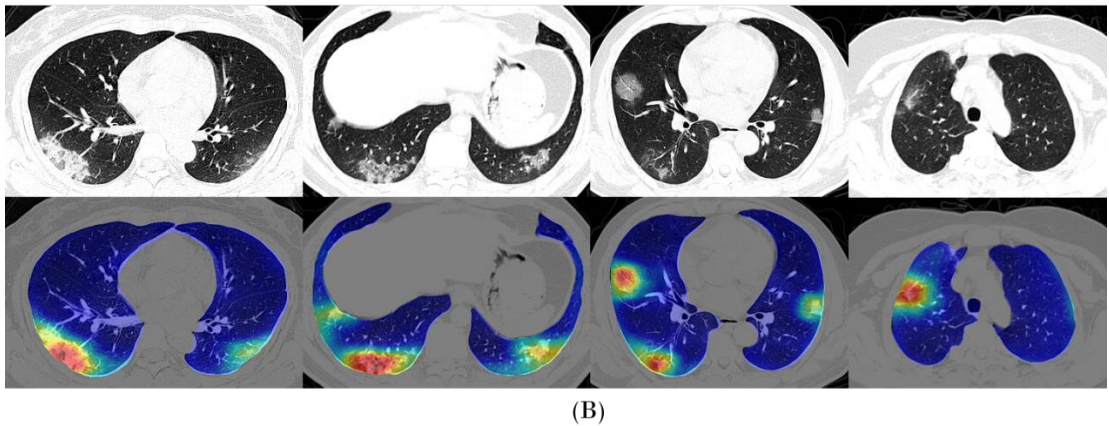

**Figure 2A:** Subsystem A: Focal opacities detection and measurements.

**Figure 2B:** Subsystem B: Slices classified as positive for Coronavirus abnormalities and their corresponding "heatmap".

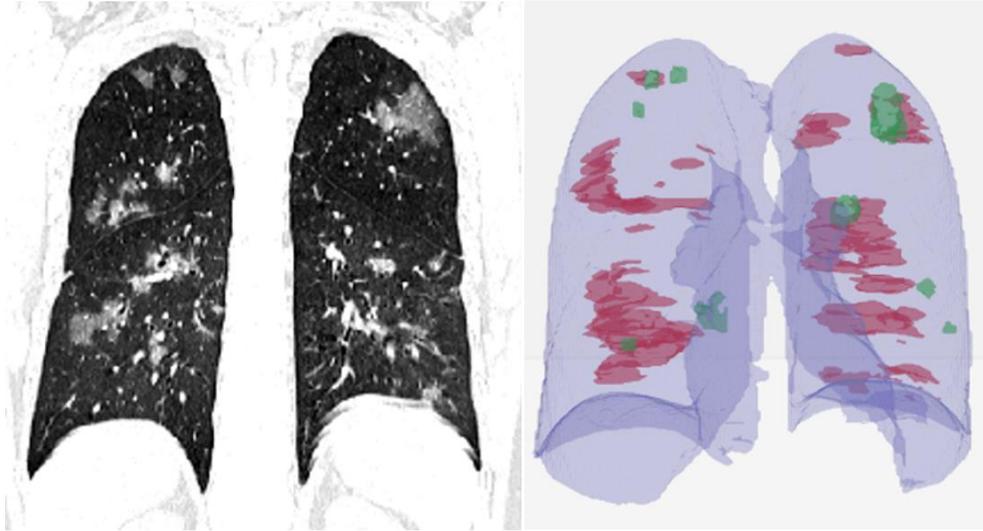

**Figure 3:** Patient case visualization. Left: Coronal view; Right: Automatically generated 3D volume map of focal opacities (green) and larger diffuse opacities (red).

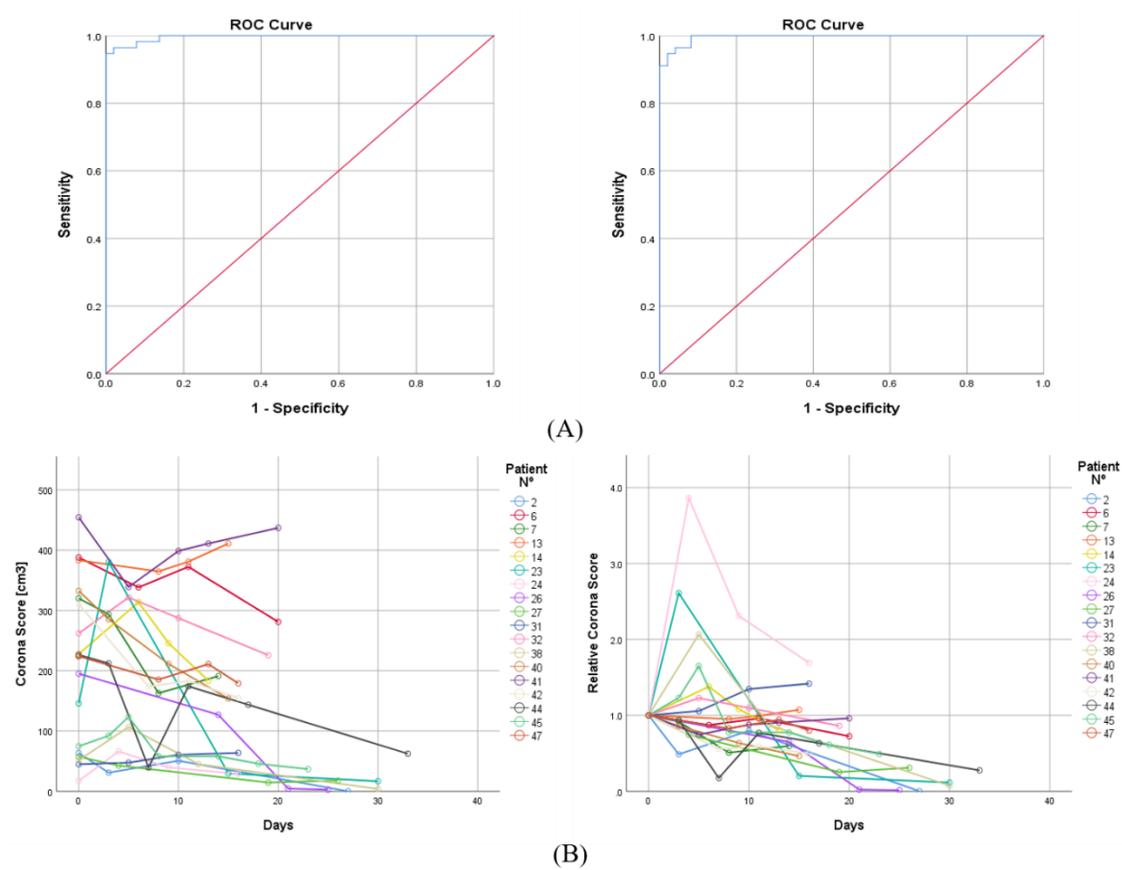

**Figure 4A**: ROC curves for the task of COVID-19 detection using *positive ratio* feature.
**Left:** Chinese Population
**Right:** Coronavirus patients from China; Non-Coronavirus patients from US

**Figure 4B**: Tracking of patient's disease progression over time using *Corona Score* (Left) and *Relative Corona Score* (Right). Day 0 corresponds to 1-4 days following first signs of the virus.

| CT Scan #1 - 26 Jan 2020 | CT Scan #2 - 30 Jan 2020 | CT Scan #3 - 4 Feb 2020 | CT Scan #4 - 11 Feb 2020 |
| --- | --- | --- | --- |
| 1-4 days after symptoms appeared | 5-8 days after symptoms appeared | 10-13 days after symptoms appeared | 17-20 days after symptoms appeared |
| 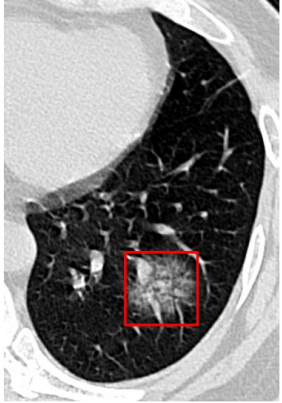 | 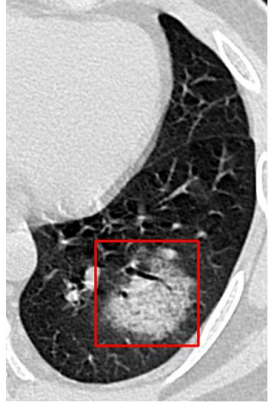 | 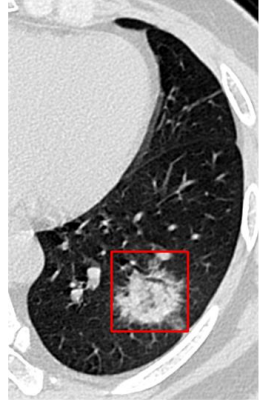 | 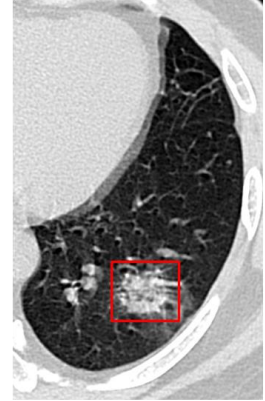 |
| Volume: 9.9 cm3 | Volume: 19.9 cm3 | Volume: 13.2 cm3 | Volume: 6.3 cm3 |
| Average Axial Diameter: 30.3 mm | Average Axial Diameter: 37.4 mm | Average Axial Diameter: 34.2 mm | Average Axial Diameter: 26.6 mm |

**Figure 5**: Multi time point tracking of patient disease progression

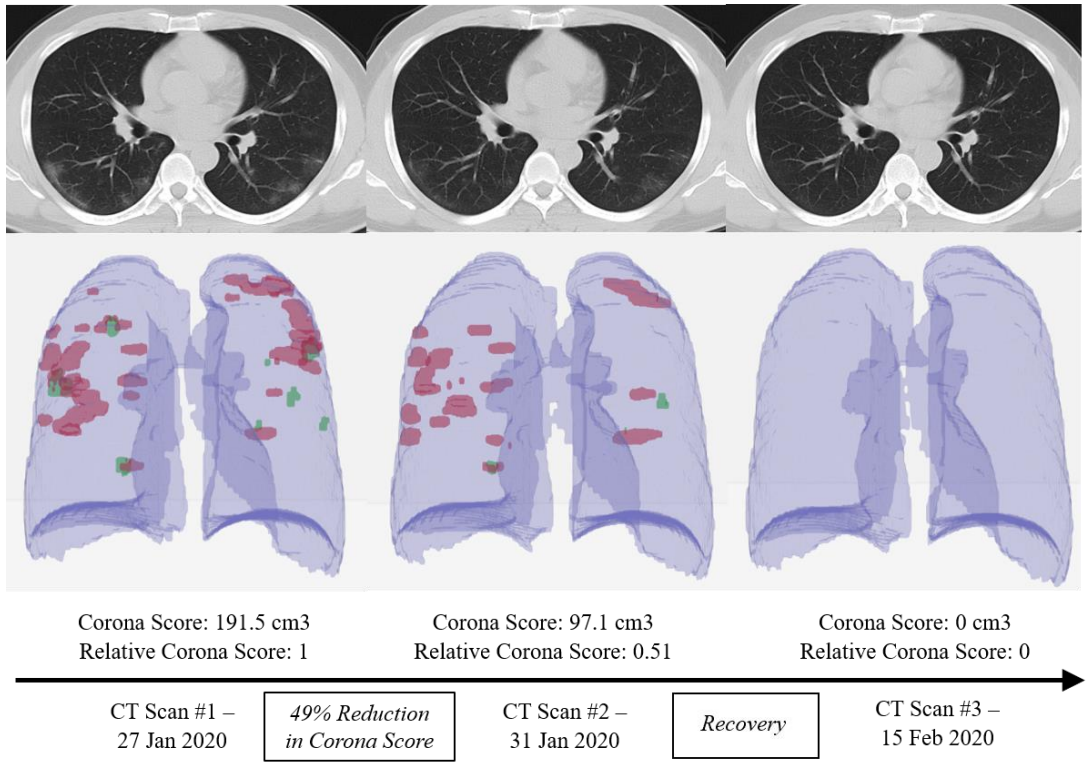

**Figure 6:** Corona score for patient disease progression monitoring